\documentclass[prl,twocolumn,showpacs,preprintnumbers,superscriptaddress,nofootinbib]{revtex4}
\usepackage{amsmath}
\usepackage{physymb}
\usepackage{siunitx}

\usepackage{tikz}
\usepackage{pgfplots}
\usepackage{pgfplotstable}
\usetikzlibrary{external}
\usetikzlibrary{matrix}
\usetikzlibrary{patterns}
\usepgfplotslibrary{groupplots}

\tikzsetexternalprefix{plotfiles/}
\pgfplotsset{compat=newest}

\pgfplotsset{
 result axis/.style={
  ymin=1e-6, 
  xlabel={$p_\perp [\si{GeV}]$},
  ylabel={$\frac{\udddc N}{\udc \eta\uddc p_\perp} \bigl[\si{GeV^{-2}}\bigr]$},
  legend style={cells={anchor=west}},
  legend columns=1,
  y filter/.code={\ifx\pgfmathresult\empty\def\pgfmathresult{-40}\fi}
 },
 data plot/.style={
  black,line width={0.05pt},mark=*,mark size={0.9pt},error bars/y dir=both,error bars/y explicit
 }
}

\pgfplotsset{
 LO/.style={
  /pgfplots/result plot fill/.style={
   /tikz/draw=none,
   /tikz/preaction={
    /tikz/fill=#1,
    /tikz/fill opacity=0.2
   },
   /tikz/pattern=crosshatch,
   /tikz/pattern color=#1,
   /pgfplots/area legend
  },
  /pgfplots/result plot line/.style={
   /tikz/draw=#1,
   /tikz/mark=x,
   /tikz/mark size={0.5pt},
   /tikz/mark options={fill=#1},
   /pgfplots/forget plot
  }
 },
 NLO/.style={
  /pgfplots/result plot fill/.style={
   /tikz/draw=none,
   /tikz/fill=#1,
   /tikz/fill opacity=0.6,
   /pgfplots/area legend
  },
  /pgfplots/result plot line/.style={
   /tikz/draw=#1,
   /tikz/mark=x,
   /tikz/mark size={0.5pt},
   /tikz/mark options={fill=#1},
   /pgfplots/forget plot
  }
 },
 rcBK LO/.style={
  LO=cyan!50!blue
 },
 rcBK NLO/.style={
  NLO=red
 },
 BK LO/.style={
  LO=blue
 },
 BK NLO/.style={
  NLO=magenta
 },
 MV LO/.style={
  LO=green!50!black
 },
 MV NLO/.style={
  NLO=orange
 },
 GBW LO/.style={
  LO=green
 },
 GBW NLO/.style={
  NLO=brown
 }
}

\newcommand\resultplot[6]{
 \addplot[#1,result plot fill] table[x expr=#2,y expr=#3] {#6};
 \addplot[#1,result plot line] table[x expr=#2,y expr=#3] {#4};
 \addplot[#1,result plot line] table[x expr=#2,y expr=#3] {#5};
}

\newcommand\setplotshift[2]{\expandafter\newcommand\csname plotshift-#1\endcsname{#2}}
\newcommand\plotshift[1]{\csname plotshift-#1\endcsname}

\newcommand\hadronconversionfactor{1.3}

\newcommand\fixedcoupling[1]{1}
\newcommand\runningcoupling[1]{pi/(2.25*(ln(greater(#1,1)*#1+notgreater(#1,1)*1) - ln(0.0588)))/0.2}
\newcommand\coupling[1]{\runningcoupling{#1}}

\tikzifexternalizing{
 \pgfplotstablesort[row sep=\\]\brahmsdAuloY{
 pt          ptlo                pthi                yield       staterr         syserr         \\
 0.65        0.6                 0.7                 0.388093    0.00243152      0.0582139      \\
 0.75        0.7                 0.8                 0.241047    0.00225041      0.036157       \\
 0.85        0.8                 0.9                 0.165855    0.000938941     0.0248783      \\
 0.95        0.9                 1                   0.112226    0.00107049      0.0168339      \\
 1.09        1                   1.2                 0.0611503   0.000486684     0.00917254     \\
 1.29        1.2                 1.4                 0.0299976   0.000307453     0.00449965     \\
 1.49        1.4                 1.6                 0.0136606   0.000190933     0.00204909     \\
 1.69        1.6                 1.8                 0.0068023   0.000125613     0.00102034     \\
 1.89        1.8                 2                   0.00399323  9.32567e-05     0.000598985    \\
 2.13        2.0                 2.3                 0.00205668  5.16416e-05     0.000308502    \\
 2.55        2.3                 2.9                 0.000565847 9.04178e-06     8.48771e-05    \\
 3.33        2.9                 4.0                 9.39018e-05 6.84736e-06     1.40853e-05    \\
 }
 \pgfplotstablesort[row sep=\\]\brahmsdAuhiY{
 pt          ptlo                pthi                yield       staterr         syserr         \\
 0.25        0.2                 0.3                 2.01147     0.0258206       0.3017205      \\
 0.35        0.3                 0.4                 1.45906     0.0174999       0.218859       \\
 0.45        0.4                 0.5                 0.933742    0.0114546       0.1400613      \\
 0.55        0.5                 0.6                 0.466025    0.00760817      0.06990375     \\
 0.65        0.6                 0.7                 0.293151    0.00555747      0.04397265     \\
 0.75        0.7                 0.8                 0.142542    0.00358952      0.0213813      \\
 0.85        0.8                 0.9                 0.085915    0.00174568      0.01288725     \\
 0.95        0.9                 1                   0.0559974   0.00123521      0.00839961     \\
 1.09        1                   1.2                 0.0233523   0.000140574     0.00467046     \\
 1.29        1.2                 1.4                 0.0104938   8.29051e-05     0.00209876     \\
 1.49        1.4                 1.6                 0.00506653  4.97884e-05     0.001013306    \\
 1.69        1.6                 1.8                 0.00246804  3.15367e-05     0.000493608    \\
 1.89        1.8                 2                   0.00120714  1.93553e-05     0.000241428    \\
 2.13        2                   2.3                 0.000459623 9.66696e-06     6.89434e-05    \\
 2.55        2.3                 2.9                 0.000124681 3.42328e-06     1.870215e-05   \\
 3.31        2.9                 4                   1.22812e-05 7.80291e-07     1.84218e-06    \\
 }
 \newcommand\starsigmainel{2.21e6} 
 \pgfplotstablesort[row sep=\\]\stardAu{
  epilo          epihi           epi         pt          yield       staterr     syserr \\
  25             30              27.1        1.03        9900        100         1500   \\
  30             35              32.2        1.21        2750        40          290    \\
  35             40              37.2        1.38        970         20          120    \\
  40             45              42.1        1.55        315         11          48     \\
  45             50              47.1        1.72        110         6           25     \\
  50             55              52.0        1.91        46          4           10     \\
 }
 \pgfplotstablesort[sort cmp={float <}]\brahmshimurcBKloY{plotfiles/brahms-dAu-mu50-rcBKL01.Y22.summary.dat}
 \pgfplotstablesort[sort cmp={float >}]\brahmslomurcBKloY{plotfiles/brahms-dAu-mu10-rcBKL01.Y22.summary.dat}
 \pgfplotstablevertcat{\brahmsfilledrcBKloY}{\brahmshimurcBKloY}
 \pgfplotstablevertcat{\brahmsfilledrcBKloY}{\brahmslomurcBKloY}
 \pgfplotstablesort[sort cmp={float <}]\brahmshimurcBKhiY{plotfiles/brahms-dAu-mu50-rcBKL01.Y32.summary.dat}
 \pgfplotstablesort[sort cmp={float >}]\brahmslomurcBKhiY{plotfiles/brahms-dAu-mu10-rcBKL01.Y32.summary.dat}
 \pgfplotstablevertcat{\brahmsfilledrcBKhiY}{\brahmshimurcBKhiY}
 \pgfplotstablevertcat{\brahmsfilledrcBKhiY}{\brahmslomurcBKhiY}
 \pgfplotstablesort[sort cmp={float <}]\starhimurcBK{plotfiles/star-dAu-mu50-rcBKL01.Y4.summary.dat}
 \pgfplotstablesort[sort cmp={float >}]\starlomurcBK{plotfiles/star-dAu-mu10-rcBKL01.Y4.summary.dat}
 \pgfplotstablevertcat{\starfilledrcBK}{\starhimurcBK}
 \pgfplotstablevertcat{\starfilledrcBK}{\starlomurcBK}
 \pgfplotstablesort[sort cmp={float <}]\brahmshimuGBWhiY{plotfiles/brahms-dAu-mu50-GBW.Y32.summary.dat}
 \pgfplotstablesort[sort cmp={float >}]\brahmslomuGBWhiY{plotfiles/brahms-dAu-mu10-GBW.Y32.summary.dat}
 \pgfplotstablevertcat{\brahmsfilledGBWhiY}{\brahmshimuGBWhiY}
 \pgfplotstablevertcat{\brahmsfilledGBWhiY}{\brahmslomuGBWhiY}
 \pgfplotstablesort[sort cmp={float <}]\brahmshimuMVhiY{plotfiles/brahms-dAu-mu50-MV24.Y32.summary.dat}
 \pgfplotstablesort[sort cmp={float >}]\brahmslomuMVhiY{plotfiles/brahms-dAu-mu10-MV24.Y32.summary.dat}
 \pgfplotstablevertcat{\brahmsfilledMVhiY}{\brahmshimuMVhiY}
 \pgfplotstablevertcat{\brahmsfilledMVhiY}{\brahmslomuMVhiY}
 \pgfplotstablesort[sort cmp={float <}]\brahmshimuBKhiY{plotfiles/brahms-dAu-mu50-BK10.Y32.summary.dat}
 \pgfplotstablesort[sort cmp={float >}]\brahmslomuBKhiY{plotfiles/brahms-dAu-mu10-BK10.Y32.summary.dat}
 \pgfplotstablevertcat{\brahmsfilledBKhiY}{\brahmshimuBKhiY}
 \pgfplotstablevertcat{\brahmsfilledBKhiY}{\brahmslomuBKhiY}
 \pgfplotstablesort[sort cmp={float <}]\totemhimurcBKhiY{plotfiles/totem-pPb-mu100-rcBKL01.Y591.summary.dat}
 \pgfplotstablesort[sort cmp={float >}]\totemlomurcBKhiY{plotfiles/totem-pPb-mu20-rcBKL01.Y591.summary.dat}
 \pgfplotstablevertcat{\totemfilledrcBKhiY}{\totemhimurcBKhiY}
 \pgfplotstablevertcat{\totemfilledrcBKhiY}{\totemlomurcBKhiY}
 \pgfplotstablesort[sort cmp={float <}]\lhcfhimurcBK{plotfiles/lhcf-pPb-mu10-rcBKL01.Y83.summary.dat}
 \pgfplotstablesort[sort cmp={float >}]\lhcflomurcBK{plotfiles/lhcf-pPb-mu02-rcBKL01.Y83.summary.dat}
 \pgfplotstablevertcat{\lhcffilledrcBK}{\lhcfhimurcBK}
 \pgfplotstablevertcat{\lhcffilledrcBK}{\lhcflomurcBK}
 \pgfplotstablesort[sort cmp={float <},sort key={mu2factor}]\rhicvsmurcBK{plotfiles/rhicvsmu-dAu-rcBKL01.Y4.summary.dat}
 \pgfplotstablesort[sort cmp={float <},sort key={mu2factor}]\lhcvsmurcBKhiY{plotfiles/lhcvsmu-pPb-rcBKL01.Y591.summary.dat}
}{} 

\newcommand\pA{\ensuremath{\text{pA}}}
\newcommand\dA{\ensuremath{\text{dA}}}

\newcommand\dAu{\ensuremath{\text{dAu}}}
\newcommand\pPb{\ensuremath{\text{pPb}}}

\begin{document}

\title{Towards the Test of Saturation Physics Beyond Leading Logarithm}

\author{Anna M. Sta\'sto}
\affiliation{Department of Physics, Pennsylvania State University, University Park, PA 16802, USA\\
RIKEN center, Brookhaven National Laboratory, Upton,  NY 11973, USA\\
Institute of Nuclear Physics, Polish Academy of Sciences, ul. Radzikowskiego 152, Krak\'ow, Poland}

\author{Bo-Wen Xiao}
\affiliation{Key Laboratory of Quark and Lepton Physics (MOE) and Institute
of Particle Physics, Central China Normal University, Wuhan 430079, China}

\author{David Zaslavsky}
\affiliation{Department of Physics, Pennsylvania State University, University Park, PA 16802, USA}

\begin{abstract}
We present results from the first next-to-leading order (NLO) numerical analysis of forward hadron production in \pA{}  and \dA{} collisions in the small-$x$ saturation formalism. Using parton distributions and fragmentation functions at NLO, as well as the dipole amplitude from the solution to the Balitsky-Kovchegov equation with running coupling, together with the NLO corrections to the hard coefficients, we obtain a good description of the available RHIC data in \dAu{} collisions. We also comment on the results in the large $p_\perp$ region beyond the saturation scale.
Furthermore, we make predictions for forward hadron production in \pPb{} collisions at the LHC. This analysis not only incorporates the important NLO corrections for all partonic channels, but also reduces the renormalization scale dependence and therefore helps to significantly reduce the theoretical uncertainties. It therefore provides a precise test of saturation physics beyond the leading logarithmic approximation.
\end{abstract}
\pacs{24.85.+p, 12.38.Bx, 12.39.St}
\maketitle

{\it 1. Introduction.} 
Prior to the era of quantum chromodynamics (QCD), the study of the physics of strong interactions at high energy was mostly based on the analytic properties of the scattering matrix. Hadron scattering was described in terms of Reggeon and Pomeron exchanges, with the latter being dominant at high center-of-mass energies. Shortly after the discovery of the QCD as the microscopic theory for the strong interaction, the BFKL (Balitskii-Fadin-Kuraev-Lipatov) Pomeron~\cite{Balitsky:1978ic} was derived using perturbative calculations in QCD. 
It predicted a strong rise of the gluon density with decreasing longitudinal momentum fraction $x$, which in turn implied  a strong growth of the cross section with increasing energy.
On a microscopic level it was understood that this strong growth is due to the Bremsstrahlung gluon radiation, which is enhanced in the small-$x$ regime. As a result, large logarithms $\left(\alpha_s \ln 1/x_g\right)^n$ appear and they can be resummed by solving the BFKL equation. 

Furthermore, it is expected that, when too many gluons are squeezed in a confined hadron, they start to overlap and recombine. The balance between radiation and recombination is known as gluon saturation. To include the effect of saturation, a nonlinear term in the evolution equation was introduced~\cite{Gribov:1984tu,Mueller:1985wy}. The derivation of the nonlinear evolution equation, the BK-JIMWLK (Balitskii-Kovchegov-Jalilian-Marian-Iancu-McLerran-Weigert-Leonidov-Kovner) equation, was performed in Refs~\cite{Balitsky:1995ub,JalilianMarian:1997jx,Iancu:2000hn}. A characteristic feature of the solution to this nonlinear evolution equation is the emergence of a dynamical scale, the saturation momentum $Q_s(x_g)$, which separates the dense saturated parton regime from the dilute regime. This type of dense and saturated gluonic matter is also known as the color glass condensate~\cite{McLerran:1993ni}. 

The quest for an experimental signal of gluon saturation has been especially important in the context of the nucleon-nucleus experiments at RHIC and the LHC~\cite{Albacete:2013ei}, and constitutes a vital part of the scientific program for possible future electron-ion colliders~\cite{eic}. There has been a lot of experimental effort devoted to the test of saturation physics both at RHIC \cite{Arsene:2004ux, Adams:2006uz, Braidot:2010ig, Adare:2011sc} and the LHC \cite{ALICE:2012xs, ALICE:2012mj, Hadjidakis:2011zz, LHCb}. Among many experimental observables which can reveal the parton saturation phenomenon, forward single inclusive hadron productions in proton-nucleus (\pA) collisions is unique in terms of its simplicity and accuracy. 

Forward single inclusive hadron production in \pA{} collisions $p+A\to h(y,p_{\perp})+X$ at leading order (LO) can be viewed as follows: a collinear parton, with momentum fraction $x$, from the proton projectile scatters off the dense nuclear target $A$ and subsequently fragments into a hadron with momentum fraction $z$ which is measured at forward rapidity $y$ with transverse momentum $p_\perp$. Observing the produced hadron at forward rapidity $y$ is particularly interesting since the proton projectile, with relatively large $x\equiv \frac{p_\perp }{z\sqrt s}e^{\eta}$, is always dilute (when $x>0.1$) while the nuclear target, with small $x_g\equiv \frac{p_\perp }{z\sqrt s}e^{-\eta}$ (when $x_g\ll 10^{-2}$), is dense in this kinematic region. When $p_\perp \leq Q_s(x_g)$, with $Q_s(x_g) \gg \Lambda_{\textrm{QCD}}$ ($Q_s$ increases as $x_g$ decreases), one should expect that gluon saturation plays an important role, while the traditional collinear factorization, which does not include multiple scatterings and small-$x$ evolutions, breaks down. This observable is also relatively simple in the large $N_c$ limit since it only depends on the dipole amplitude, which has been studied most extensively. Previous phenomenological studies~\cite{Albacete:2013ei, Dumitru:2002qt, Dumitru:2005kb, Albacete:2010bs, Levin:2010dw, Fujii:2011fh, Albacete:2012xq} on this topic either used the LO effective factorization~\cite{Dumitru:2002qt} or the $k_t$ factorization together with the running coupling corrections to the leading logarithmic (LL) BK equation. However, to test the saturation physics predictions more rigorously, one needs to go beyond the LO formalism. 

In the past few years, there has been considerable progress in the development of the effective small-$x$ factorization~\cite{Dominguez:2010xd, Altinoluk:2011qy, Chirilli:2011km} for high energy scattering in dilute-dense systems. In particular, the cross section for single inclusive hadron production in \pA{} collisions has been computed up to the one-loop order for all partonic channels~\cite{Chirilli:2011km}. It was demonstrated that the collinear and rapidity divergencies can be systematically accommodated into the parton densities, fragmentation functions and dipole amplitude, and that  the effective factorization holds up to next-to-leading order. To include the important NLO corrections, we have developed a program called Saturation physics at One Loop Order, or SOLO. The objective of this Letter is to use SOLO to numerically study single inclusive forward hadron production for the first time up to NLO accuracy, which shall shed light on the study of the onset of gluon saturation at RHIC and the LHC.

{\it 2. The numerical study of single inclusive hadron productions in proton(deuteron)-nucleus collisions.} 
We use the one-loop order results from Ref.~\cite{Chirilli:2011km} for the single inclusive cross section in \pA{} collisions calculated in the effective small-$x$ factorization formalism. The result can be schematically written as follows:
\begin{multline}
\udc\sigma =\int xf_a(x)\otimes D_a(z) \otimes \mathcal{F}^{x_g}_a(k_\perp) \otimes \mathcal{H}^{(0)} \\
+\frac{\alpha_s }{2\pi} \int xf_a(x)\otimes D_b(z) \otimes \mathcal{F}_{(N)ab}^{x_g}\otimes \mathcal{H}_{ab}^{(1)},
\label{eq:master}
\end{multline}
with the full expression available in Ref.~\cite{Chirilli:2011km}.  The sign $\otimes$ indicates the convolution between parton distributions, fragmentation functions, dipole amplitudes and hard coefficients in terms of fractions of momenta and coordinates. Here, the first line comes from the leading order contribution, while the second line comes from the one-loop calculation with the collinear divergences absorbed into the proton's collinear parton distribution $xf(x)$ or the hadron fragmentation function $D(z)$ and the rapidity divergence absorbed into the small-$x$ dipole gluon distribution $\mathcal{F}^{x_g}(k_\perp)\equiv \int \frac{d^2 r}{(2\pi)^2}e^{-ik_\perp\cdot r}S_{x_g}^{(2)}(r)$. 
It is well known that $\alpha_s$, $xf(x)$ and $D(z)$ depend on the factorization scale $\mu$. Unlike the LO formalism which contains monotonic dependence on the scale $\mu$, the NLO results that we are using come with the advantage of reduced $\mu$ dependence.
All eleven hard factors $\mathcal{H}^{(1)}$ calculated from all possible partonic channels are finite and free of divergence of any kind. Due to the high gluon density inside the heavy nucleus target, we have to take into account multiple interactions as well as the nonlinear dynamics which are encoded in the nonlinear dipole amplitude $\mathcal{F}_{(N)}^{x_g}$. The challenge of the following analysis mainly comes from the incorporation of various higher order corrections and the multi-dimensional numerical integration of these terms.

At LO, there are two diagonal channels, $q\to q$ and $g\to g$. This is indicated in the LO part of Eq.~\eqref{eq:master} with $a=q,g$, respectively. At NLO, there are also off-diagonal channels which can be computed from the second line of Eq.~\eqref{eq:master}, when $a=q,b=g$ and $a=g,b=q$. All four channels have to be taken into account in the computation. We take the large $N_c$ approximation throughout the whole computation to eliminate higher point correlation amplitudes beyond the dipole amplitude, such as quadrupoles, etc. This indicates that our results could have corrections of order $\mathcal{O}(\frac{1}{N_c^2})$.
In addition, we can also simplify the non-linear amplitudes $\mathcal{F}_{(N)}^{x_g}$ as products of dipole amplitudes in the large $N_c$ limit.  

We use the NLO MSTW parton distributions~\cite{Martin:2009iq} and NLO DSS fragmentation functions~\cite{deFlorian:2007aj} together with the one-loop running coupling. In principle, to achieve the complete NLL accuracy, we should solve the NLL BK evolution equation and obtain the NLL dipole gluon distribution for $\mathcal{F}^{x_g}(k_\perp)$ as well. However, the NLL BK equation is notoriously hard to solve due to the multidimensional integrations involved. Also, from previous analysis of the linear BFKL equation at NLL level, it is known that the higher order corrections in $\alpha_s \ln 1/x$ are very large and lead to instabilities of the evolution kernel, making resummation of collinear divergencies necessary~\cite{Ciafaloni:2003rd}. 

Thus, for practical purposes, we use the solution to the LL BK equation in coordinate space with the running coupling correction from Ref.~\cite{Balitsky:2006wa}, in which all parameters are fixed by comparing with RHIC data. Then we perform a Hankel-Fourier transform to convert the solution into momentum space in order to obtain the expression for $\mathcal{F}^{x_g}(k_{\perp})$.
The running coupling BK equation (rcBK) has been solved in many works, for example~\cite{GolecBiernat:2001if,Berger:2010sh,Albacete:2007yr}, and it was found that the inclusion of the running coupling slows down the evolution significantly, compared to the pure LL with fixed coupling, and reduces the value of the saturation scale. 
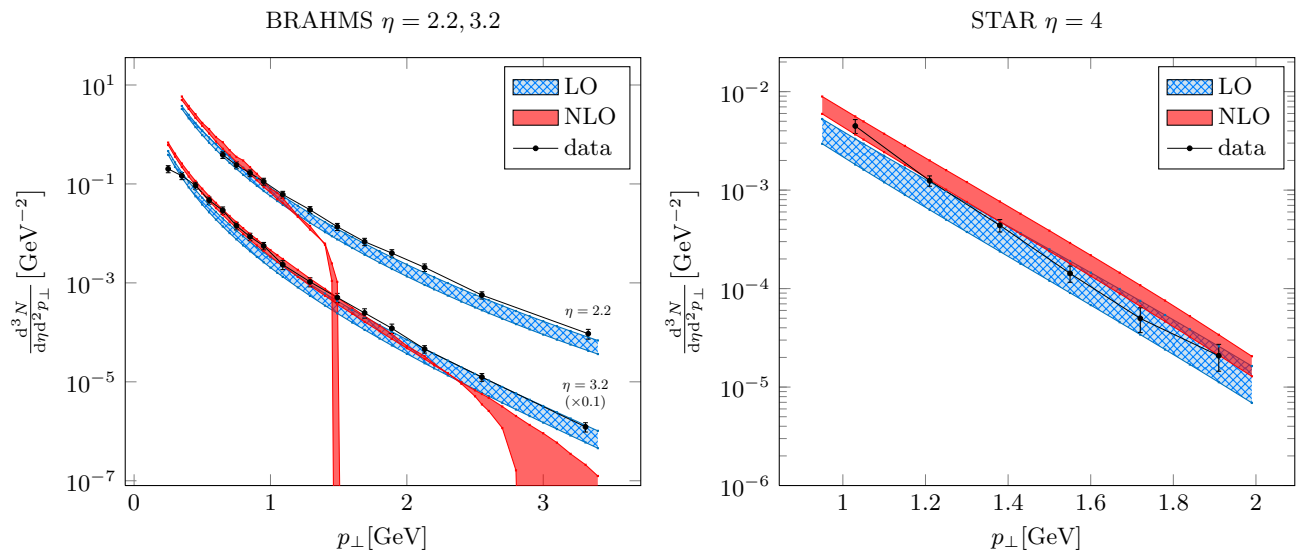
\begin{figure*}
 \setplotshift{brahms22}{1}
 \setplotshift{brahms32}{0.1}
 \tikzsetnextfilename{fig1a-BRAHMS}
 \begin{tikzpicture}
  \begin{axis}[result axis,ymode=log,ymin=8e-8,title={BRAHMS $\eta=2.2,3.2$}]
   \resultplot{rcBK LO}{\thisrow{pT}}{\plotshift{brahms32} *\hadronconversionfactor *\thisrow{lomean}}{\brahmshimurcBKhiY}{\brahmslomurcBKhiY}{\brahmsfilledrcBKhiY}
   \resultplot{rcBK NLO}{\thisrow{pT}}{\plotshift{brahms32} *\hadronconversionfactor *(\thisrow{lomean}+\coupling{\thisrow{mu2}}*\thisrow{nlomean})}{\brahmshimurcBKhiY}{\brahmslomurcBKhiY}{\brahmsfilledrcBKhiY}
   \addplot[data plot] table[x index=0,y expr={\plotshift{brahms32} *\thisrow{yield}},y error expr={\plotshift{brahms32} *(\thisrow{staterr} + \thisrow{syserr})}] {\brahmsdAuhiY}
     node[pos=1,above=1.7mm,scale=0.6,align=center] {$\eta=3.2$\\$(\times 0.1)$};
   
   \resultplot{rcBK LO}{\thisrow{pT}}{\plotshift{brahms22} *\hadronconversionfactor *\thisrow{lomean}}{\brahmshimurcBKloY}{\brahmslomurcBKloY}{\brahmsfilledrcBKloY}
   \resultplot{rcBK NLO}{\thisrow{pT}}{\plotshift{brahms22} *\hadronconversionfactor *(\thisrow{lomean}+\coupling{\thisrow{mu2}}*\thisrow{nlomean})}{\brahmshimurcBKloY}{\brahmslomurcBKloY}{\brahmsfilledrcBKloY}
   \addplot[data plot] table[x index=0,y expr={\plotshift{brahms22} *\thisrow{yield}},y error expr={\plotshift{brahms22} *(\thisrow{staterr} + \thisrow{syserr})}] {\brahmsdAuloY} node[pos=1,above=1.5mm,scale=0.6] {$\eta=2.2$};
   
   \legend{LO,NLO,data}
  \end{axis}
 \end{tikzpicture}
 \tikzsetnextfilename{fig1b-STAR}
 \begin{tikzpicture}
  \begin{axis}[result axis,ymode=log,title={STAR $\eta=4$}]
   \resultplot{rcBK LO}{\thisrow{pT}}{\thisrow{lomean}}{\starhimurcBK}{\starlomurcBK}{\starfilledrcBK}
   \resultplot{rcBK NLO}{\thisrow{pT}}{\thisrow{lomean}+\coupling{\thisrow{mu2}}*\thisrow{nlomean}}{\starhimurcBK}{\starlomurcBK}{\starfilledrcBK}
   \addplot[data plot] table[x index=3,y expr={\thisrow{yield}/\starsigmainel},y error expr={(\thisrow{staterr} + \thisrow{syserr})/\starsigmainel}] {\stardAu};
   
   \legend{LO,NLO,data}
  \end{axis}
 \end{tikzpicture}
 \caption{Comparisons of BRAHMS \cite{Arsene:2004ux} ($h^-$) and STAR \cite{Adams:2006uz} ($\pi^0$) yields in \dAu{} collisions to results of the numerical calculation with the rcBK gluon distribution, both at leading order (tree level) and with NLO corrections included. The edges of the solid bands were computed using $\mu^2 = \text{\SIrange{10}{50}{GeV^2}}$.
 }
 \label{fig:rhicresults}
\end{figure*}

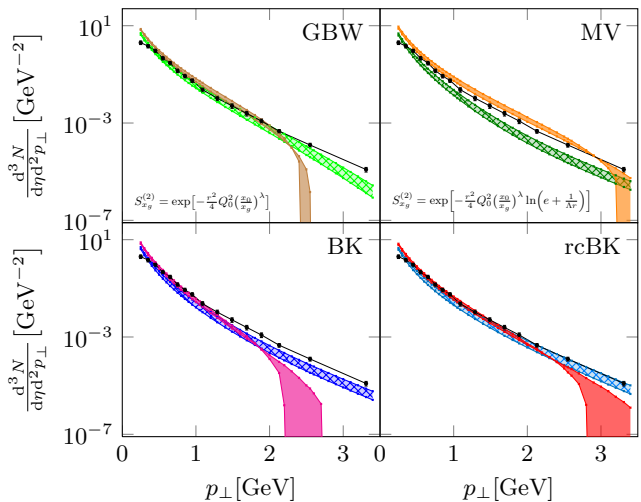
\begin{figure}[b]
 \setplotshift{brahms32}{1}
 \tikzsetnextfilename{fig2-BRAHMS}
 \begin{tikzpicture}
  \begin{groupplot}[scale=0.5,result axis,ymode=log,ymin=8e-8,ymax=5e1,xmin=0,xmax=3.5,group style={group size=2 by 2,horizontal sep=0pt,vertical sep=0pt,x descriptions at=edge bottom,y descriptions at=edge left}]
   \nextgroupplot
   \resultplot{GBW LO}{\thisrow{pT}}{\plotshift{brahms32} *\hadronconversionfactor *\thisrow{lomean}}{\brahmshimuGBWhiY}{\brahmslomuGBWhiY}{\brahmsfilledGBWhiY}
   \resultplot{GBW NLO}{\thisrow{pT}}{\plotshift{brahms32} *\hadronconversionfactor *(\thisrow{lomean}+\coupling{\thisrow{mu2}}*\thisrow{nlomean})}{\brahmshimuGBWhiY}{\brahmslomuGBWhiY}{\brahmsfilledGBWhiY}
   \addplot[data plot,mark size=0.6pt] table[x index=0,y expr={\plotshift{brahms32} *\thisrow{yield}},y error expr={\plotshift{brahms32} *(\thisrow{staterr} + \thisrow{syserr})}] {\brahmsdAuhiY};
   \node[anchor=north east] (gbwlabel) at (axis description cs:0.97,0.97) {GBW};
   \node[anchor=south west,scale=0.5] (gbwmodel) at (axis description cs:0.03,0.03) {$S^{(2)}_{x_g} = \exp\bigl[-\frac{r^2}{4} Q_0^2\bigl(\frac{x_0}{x_g}\bigr)^\lambda\bigr]$};
   
   \nextgroupplot
   \resultplot{MV LO}{\thisrow{pT}}{\plotshift{brahms32} *\hadronconversionfactor *\thisrow{lomean}}{\brahmshimuMVhiY}{\brahmslomuMVhiY}{\brahmsfilledMVhiY}
   \resultplot{MV NLO}{\thisrow{pT}}{\plotshift{brahms32} *\hadronconversionfactor *(\thisrow{lomean}+\coupling{\thisrow{mu2}}*\thisrow{nlomean})}{\brahmshimuMVhiY}{\brahmslomuMVhiY}{\brahmsfilledMVhiY}
   \addplot[data plot,mark size=0.6pt] table[x index=0,y expr={\plotshift{brahms32} *\thisrow{yield}},y error expr={\plotshift{brahms32} *(\thisrow{staterr} + \thisrow{syserr})}] {\brahmsdAuhiY};
   \node[anchor=north east] (mvlabel) at (axis description cs:0.97,0.97) {MV};
   \node[anchor=south west,scale=0.5] (mvmodel) at (axis description cs:0.03,0.03) {$S^{(2)}_{x_g} = \exp\Bigl[-\frac{r^2}{4} Q_0^2\bigl(\frac{x_0}{x_g}\bigr)^\lambda\ln\Bigl(e + \frac{1}{\Lambda r}\Bigr)\Bigr]$};
   
   \nextgroupplot
   \resultplot{BK LO}{\thisrow{pT}}{\plotshift{brahms32} *\hadronconversionfactor *\thisrow{lomean}}{\brahmshimuBKhiY}{\brahmslomuBKhiY}{\brahmsfilledBKhiY}
   \resultplot{BK NLO}{\thisrow{pT}}{\plotshift{brahms32} *\hadronconversionfactor *(\thisrow{lomean}+\coupling{\thisrow{mu2}}*\thisrow{nlomean})}{\brahmshimuBKhiY}{\brahmslomuBKhiY}{\brahmsfilledBKhiY}
   \addplot[data plot,mark size=0.6pt] table[x index=0,y expr={\plotshift{brahms32} *\thisrow{yield}},y error expr={\plotshift{brahms32} (\thisrow{staterr} + \thisrow{syserr})}] {\brahmsdAuhiY};
   \node[anchor=north east] (bklabel) at (axis description cs:0.97,0.97) {BK};
   
   \nextgroupplot
   \resultplot{rcBK LO}{\thisrow{pT}}{\plotshift{brahms32} *\hadronconversionfactor *\thisrow{lomean}}{\brahmshimurcBKhiY}{\brahmslomurcBKhiY}{\brahmsfilledrcBKhiY}
   \resultplot{rcBK NLO}{\thisrow{pT}}{\plotshift{brahms32} *\hadronconversionfactor *(\thisrow{lomean}+\coupling{\thisrow{mu2}}*\thisrow{nlomean})}{\brahmshimurcBKhiY}{\brahmslomurcBKhiY}{\brahmsfilledrcBKhiY}
   \addplot[data plot,mark size=0.6pt] table[x index=0,y expr={\plotshift{brahms32} *\thisrow{yield}},y error expr={\plotshift{brahms32} *(\thisrow{staterr} + \thisrow{syserr})}] {\brahmsdAuhiY};
   \node[anchor=north east] (rcbklabel) at (axis description cs:0.97,0.97) {rcBK};
  \end{groupplot}
 \end{tikzpicture}
 \caption{Comparisons of BRAHMS data~\cite{Arsene:2004ux} at $\eta=3.2$ with the theoretical results for four choices of gluon distribution: GBW, MV with $\Lambda=\SI{0.24}{GeV}$,  BK solution with fixed coupling at $\alpha_s = 0.1$, and rcBK with $\Lambda_\textrm{QCD}=\SI{0.1}{GeV}$. The edges of the solid bands show results for $\mu^2=\text{\SIrange{10}{50}{GeV^2}}$. As in other figures, the crosshatch fill shows LO results and the solid fill shows NLO results.}
 \label{fig:allgdist}
\end{figure}

\begin{figure*}
 \setplotshift{totem591}{1}
 \tikzsetnextfilename{fig3a-TOTEM}
 \begin{tikzpicture}
  \begin{axis}[result axis,ymode=log,ymin=8e-7,title={LHC $\eta=6.375$},restrict x to domain=0.5:4,unbounded coords=discard]
   \resultplot{rcBK LO}{\thisrow{pT}}{\plotshift{totem591} *\thisrow{lomean}}{\totemhimurcBKhiY}{\totemlomurcBKhiY}{\totemfilledrcBKhiY}
   \resultplot{rcBK NLO}{\thisrow{pT}}{\plotshift{totem591} *(\thisrow{lomean}+\coupling{\thisrow{mu2}}*\thisrow{nlomean})}{\totemhimurcBKhiY}{\totemlomurcBKhiY}{\totemfilledrcBKhiY}
   
   \legend{LO,NLO}
  \end{axis}
 \end{tikzpicture}
 \setplotshift{lhcf83}{1}
 \tikzsetnextfilename{fig3b-LHCf}
 \begin{tikzpicture}
  \begin{axis}[result axis,ymode=log,ymin=1e-7,title={LHC $\eta=8.765$},restrict x to domain=0.2:1,unbounded coords=discard]
   \resultplot{rcBK LO}{\thisrow{pT}}{\plotshift{lhcf83} *\thisrow{lomean}}{\lhcfhimurcBK}{\lhcflomurcBK}{\lhcffilledrcBK}
   \resultplot{rcBK NLO}{\thisrow{pT}}{\plotshift{lhcf83} *(\thisrow{lomean}+\coupling{\thisrow{mu2}}*\thisrow{nlomean})}{\lhcfhimurcBK}{\lhcflomurcBK}{\lhcffilledrcBK}
   
   \legend{LO,NLO}
  \end{axis}
 \end{tikzpicture}
 \caption{Predictions for the yields at the LHC energy $\sqrt{s_\text{NN}}=\SI{5.02}{TeV}$ in \pPb{} collisions, both at LO and with NLO corrections included, using the rcBK gluon distribution. On the left, we show results for $\pi^-$ yields at $\eta = 6.375$ ($Y_\text{CM} = 5.91$ in the center of mass frame) which falls in the range of pseudorapidities detected by TOTEM, and on the right, for $\pi^0$ yields at $\eta = 8.765$ ($Y_\text{CM} = 8.3$) which falls in the range detected by LHCf. The edges of the solid bands were computed using $\mu^2 = \text{\SIrange{20}{100}{GeV^2}}$ on the left and $\mu^2 = \text{\SIrange{2}{10}{GeV^2}}$ on the right.}
 \label{fig:lhcresults}
\end{figure*}
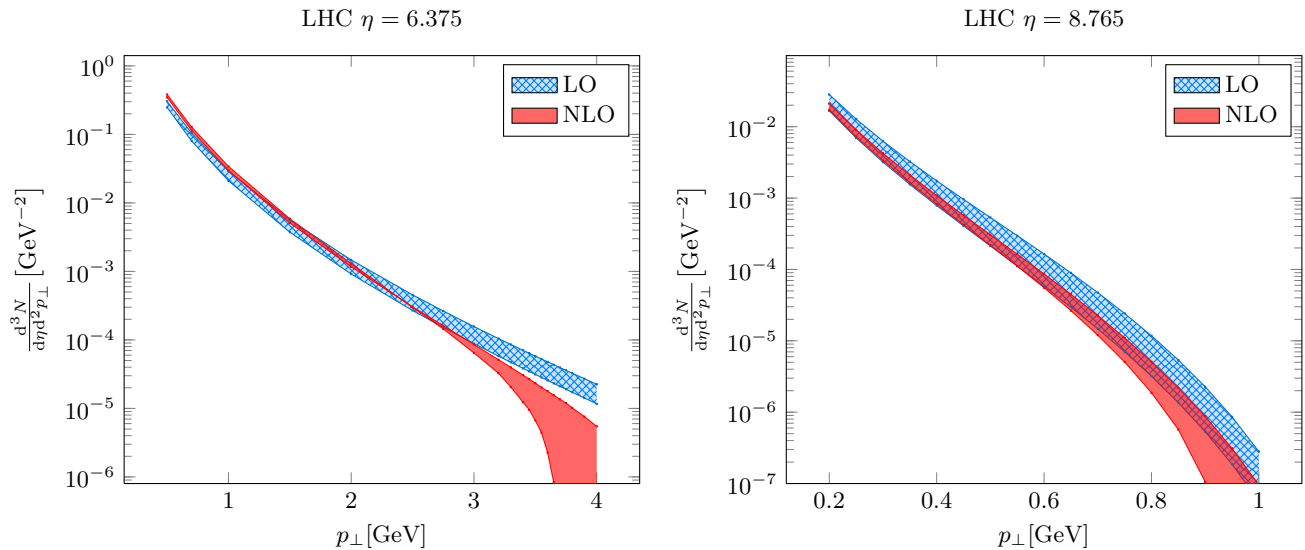

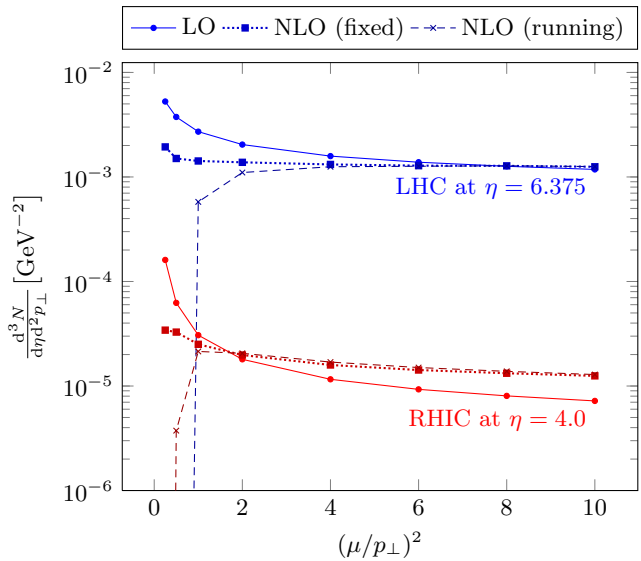
\begin{figure}
 \tikzsetnextfilename{fig4-vsmu}
 \begin{tikzpicture}
  \begin{axis}[result axis,ymode=log,xlabel=$(\mu/p_\perp)^2$,legend columns=4,every axis legend/.append style={at={(axis description cs:0.5,1.03)},anchor=south,cells={anchor=west}}]
   \addplot[blue,mark=*,mark size=1pt] table[x index=2,y expr=\thisrow{lomean}] {\lhcvsmurcBKhiY} node[pos=1,below left] {LHC at $\eta = 6.375$};
   \addplot[blue!80!black,mark=square*,mark options={solid},mark size=1pt,densely dotted,thick] table[x index=2,y expr={\thisrow{lomean}+\fixedcoupling{\thisrow{pT}^2*\thisrow{mu2factor}}*\thisrow{nlomean}}] {\lhcvsmurcBKhiY};
   \addplot[blue!60!black,mark=x,mark size=1.5pt,densely dashed] table[x index=2,y expr={\thisrow{lomean}+\runningcoupling{\thisrow{pT}^2*\thisrow{mu2factor}}*\thisrow{nlomean}}] {\lhcvsmurcBKhiY};

   \addplot[red,mark=*,mark size=1pt] table[x index=2,y expr=\thisrow{lomean}] {\rhicvsmurcBK} node[pos=1,below left] {RHIC at $\eta = 4.0$};
   \addplot[red!80!black,mark=square*,mark options={solid},mark size=1pt,densely dotted,thick] table[x index=2,y expr={\thisrow{lomean}+\fixedcoupling{\thisrow{pT}^2*\thisrow{mu2factor}}*\thisrow{nlomean}}] {\rhicvsmurcBK};
   \addplot[red!60!black,mark=x,mark size=1.5pt,densely dashed] table[x index=2,y expr={\thisrow{lomean}+\runningcoupling{\thisrow{pT}^2*\thisrow{mu2factor}}*\thisrow{nlomean}}] {\rhicvsmurcBK};
   
   \legend{LO,NLO (fixed),NLO (running)};
  \end{axis}
 \end{tikzpicture}
 \caption{$\mu$-dependence of the calculated cross sections at $p_\perp = \SI{2}{GeV}$. The NLO results for fixed coupling ($\alpha_s=0.2$) and one-loop running coupling are both presented, where the $\alpha_s$ here is referred to the one in front of the NLO hard coefficients in Eq.~(\ref{eq:master}). The dramatic drop of the NLO curve with the running coupling at low $\mu^2$ is simply due to the breakdown of perturbative calculations at large $\alpha_s (\mu)$. Nevertheless, these two NLO curves almost overlap with each other at large values of $\mu^2$.
 }
 \label{fig:vsmu}
\end{figure}

{\it 3. Results.}
Fig.~\ref{fig:rhicresults} shows the single inclusive hadron production yields in \dAu{} collisions measured by BRAHMS \cite{Arsene:2004ux} and STAR \cite{Adams:2006uz} and the corresponding curves generated by SOLO using the rcBK solution as the gluon distribution. Without use of K factors, we find generally decent agreement between our NLO calculation and the data for relatively low momenta $p_\perp <Q_s(x_g)$. The limit $Q_s(x_g)$ increases with forward rapidity $\eta$. Therefore, this calculation becomes more robust as $\eta$ increases. 

A notable feature of the present calculation is that the NLO correction becomes negative at higher $p_\perp$, and in fact dominates over the leading order result for some values of $p_\perp$. Similar behavior is also seen in Ref.~\cite{Albacete:2012xq} which incorporated only part of the NLO corrections. 
The critical value of $p_\perp$ at which the overall LO+NLO cross section becomes negative increases with rapidity, as can be seen from Fig.~\ref{fig:rhicresults}. Once the hadron transverse momentum $p_\perp$ is larger than $Q_s(x_g)$, the NLO correction starts to become very large and negative. This indicates that we need to either go beyond NLO or perform some sort of resummation when $p_\perp >Q_s(x_g)$, due to this theoretical limitation of the dilute-dense factorization formalism at NLO.
This is an important problem but it lies outside the scope of the current work and we will leave this to future study. Given these limitations, we expect the dilute-dense factorization formalism to work much better for more forward rapidity regions. This trend is indeed observed in Fig.~\ref{fig:rhicresults} and Fig.~\ref{fig:lhcresults}. Nevertheless, as shown in all the plots, the results computed from SOLO are stable and reliable as long as $p_\perp <Q_s (x_g)$.

Furthermore, we have also run SOLO with three other choices of dipole gluon distribution: the Golec-Biernat and Wusthoff (GBW) model \cite{GolecBiernat:1998js}, the McLerran-Venugopalan (MV) model \cite{McLerran:1993ni}, and the solution to the fixed coupling BK equation. As shown in Fig.~\ref{fig:allgdist}, all four parametrizations give similar results and agree with the BRAHMS data in the $p_\perp < Q_s$ region. For other plots, we only use the rcBK solution, which is the most sophisticated parametrization.

Fig.~\ref{fig:lhcresults} shows predictions made by SOLO for \pPb{} collisions at high pseudorapidities which are accessible at LHC detectors, in particular $5.3 \leq \eta \leq 6.5$ for TOTEM's T2 telescope~\cite{Berardi:2004ku} and $\eta \geq 8.4$ at LHCf~\cite{Adriani:2008zz}.
Of course, our prediction in the left plot should only be valid when $p_\perp < \SI{3}{GeV}$, which is about the size of the saturation momentum at the corresponding rapidity.

One of the advantages of the NLO results is the significantly reduced scale dependence as shown in Fig.~\ref{fig:vsmu}. In principle, cross sections for any physical observable, if it could be calculated up to all order, should be completely independent of the factorization scale $\mu$. However, as shown in Fig.~\ref{fig:vsmu}, the LO cross section is a monotonically decreasing function of the factorization scale $\mu$. This is well-known and is simply due to the fact that an increase of $\mu$ causes both the parton distribution function (in the region $x>0.1$) and the fragmentation function (in the region $z>0.2$) to decrease. Therefore, one has to choose the scale $\mu$ properly for LO calculations. By including the NLO corrections, which cancels all the scale dependence up to one-loop order, we find that the dependence on $\mu$ is sharply reduced in the NLO cross section except for very low $\mu^2$ values. In other words, the factorization scale can be chosen from a large range of values without changing the cross sections much. This greatly increases the reliability of our calculation and reduces the uncertainty of our prediction. In addition, Fig.~\ref{fig:vsmu} indicates that the best choice of factorization scale $\mu$ should be about two or maybe three times the average transverse momentum of the produced hadron. This helps us to choose a reasonable range of $\mu^2$ to set the error band for our numerical analysis. 

{\it 4. Discussion and Conclusion.}
As an important first step towards the NLO phenomenology in the saturation physics, we have developed a program called SOLO which allows us to incorporate most of the NLO corrections for forward single hadron productions in \pA{} collisions. 
We have used recent theoretical results for forward hadron production at NLO accuracy, which demonstrate the factorization of collinear and rapidity divergencies, together with NLO parton distribution functions and fragmentation functions, as well as the solution to the BK equation with running coupling.
We obtained decent agreement with the experimental data from RHIC and we have made predictions for the forward production in \pA{} collisions at the LHC. The results show the enhancement of the NLO calculation over the LO calculation at very low values of $p_\perp$, and the reduction of the NLO cross section with respect to the LO calculation at higher values of the hadron transverse momentum.  

We found that the scale dependence is significantly reduced  at NLO as compared to the lowest order result.
We also found that the results turn negative for higher values of $p_T$ above some critical value. This critical value increases with rising rapidity, thus justifying the calculation for the forward region.
Several extensions of this work are possible. The large negative value of the NLO correction may imply the need to include higher order corrections or some resummation in order to stabilize the result beyond the critical value of $p_\perp$.
Also, for the complete NLL analysis one would need to evaluate the dipole gluon distribution using the NLL BK equation. These are important issues that certainly deserve separate studies. 

Nevertheless, this calculation is important progress in small-$x$ physics phenomenology beyond LL accuracy, and it provides predictions for \pA{} collisions at the LHC with the theoretical uncertainty under control.

We thank C. Marquet, A. Mueller, S. Munier, B. Pire, L. Szymanowski and F. Yuan for their stimulating comments. This work was supported in part  by the DOE OJI grant No. DE - SC0002145   and by the Polish NCN grant DEC-2011/01/B/ST2/03915.  A.M.S. is supported by the Sloan Foundation.

\end{document}